\documentclass[aps,prb,preprint,psfig,groupedaddress]{revtex4}
\usepackage{graphicx}
\usepackage{bm}

\newcommand{\BS}{Bi$_2$Se$_3$}

\newcommand{\MBS}{Mn-Bi$_2$Se$_3$}

\renewcommand{\vec}[1]{\mbox{\boldmath$1$}}

\newcounter{lastnote}

\def\bc{\begin{center}}
\def\ec{\end{center}}
\def\be{\begin{equation}}
\def\ee{\end{equation}}
\renewcommand{\vec}[1]{\mbox{\boldmath$1$}}

\usepackage{amsmath,amssymb}

\begin{document}
\title{Interplay between ferromagnetism, surface states, and quantum corrections in a magnetically doped topological insulator}
\author{Duming Zhang$^1$, Anthony Richardella$^1$, David W. Rench$^1$, Su-Yang Xu$^2$, Abhinav Kandala$^1$, Thomas C. Flanagan$^1$, Haim Beidenkopf$^2$, Andrew L. Yeats$^3$, Bob B. Buckley$^3$, Paul V. Klimov$^3$, David D. Awschalom$^3$, Ali Yazdani$^2$, Peter Schiffer$^1$, M. Zahid Hasan$^2$ and Nitin Samarth$^1$}
\email{nsamarth@psu.edu}
\affiliation{$^1$Department of Physics and Materials Research Institute, The Pennsylvania State University, University Park, Pennsylvania 16802-6300, USA}
\affiliation{$^2$Department of Physics, Princeton University, Princeton, New Jersey 08544, USA}
\affiliation{$^3$Center for Spintronics and Quantum Computation, University of California, Santa Barbara, California 93106, USA}

\date{\today}
\begin{abstract}
The breaking of time-reversal symmetry by ferromagnetism is predicted to yield profound changes to the electronic surface states of a topological insulator. Here, we report on a concerted set of structural, magnetic, electrical and spectroscopic measurements of \MBS~ thin films wherein photoemission and x-ray magnetic circular dichroism studies have recently shown surface ferromagnetism in the temperature range 15 K $\leq T \leq 100$ K, accompanied by a suppressed density of surface states at the Dirac point. Secondary ion mass spectroscopy and scanning tunneling microscopy reveal an inhomogeneous distribution of Mn atoms, with a tendency to segregate towards the sample surface. Magnetometry and anisotropic magnetoresistance measurements are insensitive to the high temperature ferromagnetism seen in surface studies, revealing instead a low temperature ferromagnetic phase at $T \lesssim 5$ K. The absence of both a magneto-optical Kerr effect and anomalous Hall effect suggests that this low temperature ferromagnetism is unlikely to be a homogeneous bulk phase but likely originates in nanoscale near-surface regions of the bulk where magnetic atoms segregate during sample growth. Although the samples are not ideal, with both bulk and surface contributions to electron transport, we measure a magnetoconductance whose behavior is qualitatively consistent with predictions that the opening of a gap in the Dirac spectrum drives quantum corrections to the conductance in topological insulators from the symplectic to the orthogonal class. 
\end{abstract}
\maketitle

\section{Introduction}
The metallic Dirac cone surface states of a topological insulator (TI) are expected to be protected against small perturbations by time-reversal symmetry\cite{Moore_2010,Qi2010}. However, these surface states can be dramatically modified by breaking time-reversal symmetry through magnetic doping\cite{Qi2008,Qi2008b}. This is predicted to open a finite energy gap at the Dirac point\cite{Liu2009}, an important consideration for potential device applications of TIs. Further, the interplay between magnetism and TI surface states is predicted to yield novel phenomena of fundamental interest such as a topological magneto-electric effect\cite{Qi2008b}, a quantized anomalous Hall effect\cite{Yu2010}, and the induction of a magnetic monopole\cite{Qi2009}. These predictions have already triggered an active search for materials that provide experimental realizations of such phenomena. Angle-resolved photoemission spectroscopy (ARPES) studies have shown that magnetic impurities modify the surface states\cite{Wray2010} and may possibly open a surface state gap\cite{Chen2010}. Ferromagnetism has been demonstrated in several magnetically doped TI crystals, including bulk crystals of Mn-doped Bi$_2$Te$_3$ \cite{Hor2010} and Bi$_2$Te$_{3-y}$Se$_y$\cite{Checkelsky2012}, as well as thin films of V-doped Sb$_2$Te$_3$\cite{Zhou2005}, Cr-doped Sb$_2$Te$_3$\cite{Zhou2006,Chang2011} and Cr-doped \BS\cite{Haazen2012}. The evidence for ferromagnetism in these previous studies is primarily based upon magnetometry and magnetotransport, neither of which unfortunately provide a clear means of distinguishing between surface and bulk magnetism. However, the behavior of the anomalous Hall effect in electrically gated Mn-doped Bi$_2$Te$_{3-y}$Se$_y$ samples has been interpreted as evidence for surface ferromagnetism \cite{Checkelsky2012}. 

Recently \cite{Xu_NP}, a spin resolved ARPES and x-ray magnetic circular magnetic dichroism (XMCD) study of Mn-doped \BS~thin films clearly revealed the presence of out-of-plane surface ferromagnetism that persists up to relatively high temperature ($T \sim 100$ K) and demonstrated a hedgehog spin texture accompanying a Dirac-metal to gapped-insulator transition. The broadening of the ARPES data also suggested an inhomogeneous surface magnetization. Here, we report additional studies of these thin films down to low temperatures ($T \geq 300$ mK), focusing on concerted structural, magnetic, electrical, and spectroscopic measurements. Magnetometry and magneto-transport measurements indicate the existence of a low temperature ferromagnetic phase with a Curie temperature $T_C \leq 5.4$K and an in-plane magnetization. The use of multiple complementary structural probes in our \MBS~thin films suggests that this observed low temperature ferromagnetic phase is not a homogeneous bulk effect but likely occurs in a limited region ($\sim 5$nm) near the sample surface. We note that these macroscopic magnetization and transport measurements are likely to miss the purely surface out-of-plane and high $T_C$ ferromagnetism observed in spin resolved ARPES measurements and XMCD. We also note that a recent calculation \cite{Rosenberg} indicates that strictly surface ferromagnetism can be more robust than ferromagnetism mediated by bulk states, a scenario that is consistent with our observations. As the samples are cooled through $T_C$ of the low temperature ferromagnetic phase, the magnetoconductance of the samples goes through a distinctive transition from symplectic (or weak anti localization) to orthogonal (or weak localization) quantum corrections to transport. Since ARPES measurements show that the Fermi level in these samples is located well in the bulk conduction band states, the magnetoconductance data probe a complicated interplay between states that include the pure TI surface, confined two dimensional (2D) states created by band bending and three dimensional (3D) bulk states. Despite this complex situation, the onset of time reversal symmetry breaking ferromagnetism qualitatively yields the magnetoconductance behavior predicted for 2D Dirac cone TI surface transport\cite{Lu2011}. We note that a recent study\cite{Liu2011} of ultrathin (3 QL) samples of Cr-doped \BS~showed a similar cross-over from weak anti-localization to weak localization, but in a regime where the samples are {\it paramagnetic} and where a surface state gap is already developed at a temperature ($\sim100$ K) much higher than that implied by transport measurements ($ \lesssim10$ K). Given the development of a gap due to the hybridization of the upper and lower surfaces in these ultra thin samples, it is difficult to unambiguously disentangle this effect from the opening of an additional gap from possible breaking of time reversal symmetry. Nonetheless, it is quite possible that surface ferromagnetism also exists in those samples and could be detected using XMCD.

\section{Thin film synthesis and characterization}
The \MBS~films were synthesized by MBE using high purity elemental (5N) Mn, Bi, and Se sources.  After thermal desorption of the native oxide on the epi-ready GaAs 111A substrate, a thin ZnSe buffer layer ($\sim 8 $ nm) was first deposited at $\sim 300 ^{\circ}$C using a Se/Zn beam equivalent pressure (BEP) ratio of $\sim 2.3$ to yield a smooth surface. A \MBS~layer ($\sim 60$ nm) was then synthesized immediately on the ZnSe buffer with a Se/Bi BEP ratio of $\sim 15$. The Mn doping concentration was controlled by changing the Bi/Mn BEP ratio in a range of $\sim 8-24$. Reflection high energy electron diffraction (RHEED) was used to monitor the entire growth process. Figure 1 (a) shows a typical atomic force microscopy (AFM) image of a \MBS~thin film (BEP ratio Bi/Mn = 12.5) with a root mean square surface roughness $\sim 0.5$ nm. The triangular islands at the surface indicate screw dislocations. X-ray diffraction (XRD) shows that \MBS~has the same rhombohedral layered crystal structure as undoped \BS, but with strained lattice parameters. The observation of only (003) family of reflections indicates that the growth direction is oriented along the $c$ axis with no secondary phases such as MnBi or MnSe (Fig.\ 1 (b)). In highly doped samples, the smaller interplanar distance between (006) planes  indicates that the Mn atoms substitute the Bi atoms due to the smaller atomic radius of Mn (0.140 nm) relative to Bi (0.160 nm); while the larger interplanar distance between (003) planes indicates that the Mn atoms also intercalate between quintuple layers. As the Mn doping concentration increases, the quality of the thin films degrades and the crystal structure shifts towards the BiSe phase.

Inhomogeneous phase separation into nanoscale regions of ferromagnetism is well known in magnetically doped semiconductors\cite{DietlNM}. To examine this possibility, we measured the secondary ion mass spectrometry (SIMS) depth profile in several samples to reveal the Mn concentration distribution along the growth direction. Figure 1 (c) shows a depth profile from the Bi/Mn = 12.5 sample. The Mn doping concentration ($\sim 20 \%$ relative error) was determined using a sample with bulk Mn concentration of $x=0.043$ (Mn$_x$(\BS)$_{1-x}$) as a standard, the composition of which was obtained from Rutherford backscattering spectroscopy. These measurements show that the surface Mn concentration is almost an order of magnitude higher than that of the bulk. X-ray photoelectron spectroscopy (XPS) measurements provide further confirmation of excess Mn at the surface: the averaged Mn concentration from the top $\sim5$ nm of the Bi/Mn = 12.5 sample is $x=0.085$, about eight times higher than that of the bulk ($x=0.010$). To study the morphology of the excess Mn at the surface, we carried out scanning tunneling microscopy (STM) measurements on samples that were capped {\it in situ} with Se immediately after growth. Careful de-capping of these samples in an ultrahigh vacuum chamber at 200 $^{\circ}$C, {\it lower than the growth temperature of} 300 $^{\circ}$C, allows the acquisition of STM data that reveal substitutional Mn below the surface, individual Mn adatoms on the surface, as well as numerous small regions with a high density of clustered Mn atoms (Fig.\ 1 (d)). The clusters have typical dimensions of tens of nanometers, are generally located near atomic step edges (indicated by dark lines), and are distributed across the entire surface. Cross-sectional high-resolution transmission electron microscopy (HRTEM) provides further detailed insights into the bulk structural characteristics of the \MBS~ thin films. Figures 1 (e) and 1 (f) show a representative HRTEM image from a cross-sectional specimen of the Bi/Mn = 12.5 sample and the corresponding selected area electron diffraction pattern, respectively. The clear phase contrast image and diffraction pattern indicate that the \MBS~films are generally high quality single crystal with little disorder. Within our TEM resolution limit, there is no clear evidence for the formation of Mn clusters in the bulk. The observation of Mn clusters at the surface in STM studies and their absence in the bulk in HRTEM studies further confirm that there is excess surface Mn in these samples.

\section{Magnetometry measurements: observation of ferromagnetism}
After understanding the Mn distribution in our \MBS~thin films, we address the possibility of time reversal symmetry breaking ferromagnetic order amongst the Mn impurities. As we mentioned in the introduction, XMCD measurements have indeed revealed the presence of surface ferromagnetic order with out-of-plane magnetization up to at least $T = 100$ K. We have now carried out magnetization measurements using a superconducting quantum interference device (SQUID) magnetometer in a Quantum Design Magnetic Property Measurement System with a base temperature at 1.8 K. Figure 2 (a) shows temperature dependent magnetization of the Bi/Mn = 12.5 samples in field along both the $\langle1\bar{1}00\rangle$ direction and the $c$ axis. The temperature dependent magnetization curves were obtained using a 50 Oe measuring field after field cool with $H=9$ kOe. Fitting to the high-temperature data allows for the determination of a temperature-independent diamagnetic signal of the substrate, the constant offset of which is removed from the sample data. Fitting a line to the steepest slope of the curve allows us to extrapolate a ferromagnetic Curie temperature  $T_C\sim5.4$ K (indicated by the arrow). Based on the Mn distribution in this thin film, we estimate $\sim3.45 \mu_{\rm{B}}$ per Mn atom by assuming that the ferromagnetism originates from the top 5 nm near the surface with 8.5 a.t.\% Mn based on our XPS measurement. Figures 2 (b) and 2 (c) plot magnetization vs. field curves of the Bi/Mn = 12.5 sample at different temperatures in fields along the $\langle1\bar{1}00\rangle$ direction and the $c$ axis, respectively. The diamagnetic signal due to the substrate is subtracted off by fitting to the high-field data and correspondingly removing the diamagnetic background. The hysteresis above $T_C$ (Fig.\ 2 (c)) is attributed to defects and/or impurities in the GaAs substrate based on careful control measurements of several bare GaAs substrates. The easy axis is found to be {\it in plane} and close to the $\langle1\bar{1}00\rangle$ direction. Ferromagnetic ordering with a $T_C\sim3.1$ K is also observed in the Bi/Mn = 23.6 sample (Fig.\ 2d - 2f). In the two lightly doped samples (Bi/Mn = 23.6 and 12.5), $T_C$ increases with Mn doping, a trend that we have noticed in other samples grown in this doping range (data not shown). However, we also find that for high Mn doping, the quality of the crystal degrades and the crystal structure shifts towards BiSe instead of \BS. A higher Mn concentration also increases the carrier concentration significantly (up to $\sim 10^{20}$ cm$^{-3}$). Both effects seem to result in a  lower $T_C$. 

\section{Transport measurements: observation of ferromagnetism and cross-over from weak anti-localization to weak localization }
The ferromagnetic ordering in \MBS~thin films is further confirmed by magneto-transport measurements. To study electrical transport, Hall bar devices ($1300\:\mu\rm{m}\times400\:\mu$m) were fabricated from thin films using photolithography and wet etching. Transport measurements were carried out using a standard lock-in technique in an Oxford Heliox He-3 cryostat with a base temperature at 0.5 K and field up to 6 T as well as an Oxford Triton He-3 vector magnet system with a base temperature at 0.3 K wherein a 1 T field could be arbitrarily oriented over the entire sphere. For clarity, we focus on the sample with Bi/Mn $=12.5$, noting that similar behavior is also observed in other samples. The measurement geometry is shown in Fig.\ 3 (a): the Hall bar device lies in the $xy$ plane with the excitation current along the $x$ axis ($\langle1\bar{1}00\rangle$ direction). The orientation of the magnetic field $H$ is defined by the polar angle $\theta$ and the azimuthal angle $\varphi$. Figure 3 (b) shows the magnetoconductivity (MC) $\Delta\sigma_{xx}=\sigma_{xx}(H)-\sigma_{xx}(0)$ in units of $e^2/h$ at $T = 0.3$ K when the magnetic field is swept at different fixed angles in the $xz$ plane. The arrows indicate the field sweep directions. The longitudinal conductivity $\sigma_{xx}$ is obtained from the longitudinal and transverse resistivity measurements using the 2D tensor: $\sigma_{xx}=\frac{\rho_{xx}}{(\rho_{yx}^2+\rho_{xx}^2)}$. Note that hysteresis is observed in the MC for all field orientations, but is most pronounced when the magnetic field is at a slight angle from the $z$ axis. This angular dependence of the hysteresis in the MC is a strong indication of ferromagnetic ordering amongst the magnetic impurities, with a clear anisotropy that favors a magnetization oriented close to the $x$ axis. To further identify the critical temperature $T_c$ of ferromagnetism, we carried out temperature dependent measurements of the MC with the field oriented at the angle that showed the most pronounced hysteresis ($\theta=5^{\circ}$ and $\varphi=0$). Figure 3 (c) shows that the magnitude of the hysteresis and the magnetization switching field both decrease with increasing temperature, vanishing at $T_c\sim5.5$ K, which is consistent with the $T_c$ obtained from SQUID measurement (Fig.\ 2 (a)). Similar hysteretic behaviors were also observed in other samples (Fig.\ (4)).

The presence of ferromagnetism is also manifest through another classic signature in transport: anisotropic magnetoresistance (AMR). In these measurements, we carried out angular sweeps of a magnetic field with fixed magnitude in the $xy$ plane. In Fig.\ 3 (d), an in-plane magnetic field was applied with a constant magnitude of 0.5 T to saturate the magnetization while its orientation was continuously swept through $360^{\circ}$. The angular dependence of the magneto-resistance can be interpreted using the standard heuristic expression for AMR\cite{Pan1957},
\begin{equation}
E_x=j\rho_{\perp}+j(\rho_{\parallel}-\rho_{\perp})\cos^2\varphi,
\end{equation}
where $E_x$ is the longitudinal electric field within a single domain ferromagnetic film, $j$ is the current density, $\varphi$ is the angle between the magnetization and current density, and $\rho_{\parallel}$ and $\rho_{\perp}$ are the resistivities for current parallel and perpendicular to the magnetization. The fits (lines) using equation (1) describe our data (squares) very well. The minimum resistance occurs at $\varphi=0$ and $180^{\circ}$, indicating that $\rho_{\parallel}<\rho_{\perp}$. As the temperature increases, the angular dependence of the in-plane MR becomes weaker and finally isotropic above $\sim10$ K. The temperature dependence of AMR is qualitatively consistent with that of the MR hysteresis. The slightly higher temperature at which the AMR disappears is attributed to contributions from MR anisotropy which is present even in undoped \BS\cite{Wang2011b}.

The observation of ferromagnetism using magnetometry and magnetoresistance would usually imply the presence of an anomalous Hall effect (AHE). Surprisingly, we do not observe any obvious signatures of an AHE in any of the samples (Fig.\ 5). Furthermore, magneto-optical Kerr effect (MOKE) measurements using light of wavelength 690 nm do not reveal any observable signal (down to a precision of 50 $\mu$rad) at temperatures as low as 2 K.  From the absence of an AHE and MOKE signal, it is at first tempting to attribute the ferromagnetism in these samples to nanoscale phase-separated Mn-rich regions located on and near the surface, which is supported by the surface and structural characterization. While the presence of these clusters would lead to spin-dependent scattering contributions to the MC, they would be unlikely to contribute to an AHE or to a measurable MOKE signal. We are aware of one other hybrid micro-composite ferromagnet -- Cd$_{1-x}$Mn$_x$GeAs$_2$:MnAs -- where AHE is absent despite the clear observation of ferromagnetism \cite{Kilanski2011}. However, ongoing measurements of ferromagnetic resonance (FMR) in our \MBS~samples seem to suggest otherwise: the angular dependence of FMR shows the standard behavior seen in homogeneous ferromagnets \cite{Bardelben}. We do not currently have a definitive explanation for all these incongruous results, but a reasonable scenario is the following. The out-of-plane surface ferromagnetism seen by XMCD at high temperatures is clearly undetectable using SQUID magnetometry, MOKE and AHE. As the sample temperature is lowered below $\sim 5$ K, the Mn-rich bulk region within $\sim 5$ nm of the surface undergoes another ferromagnetic phase transition that is seen by SQUID and by FMR, as predicted by mean field theory \cite{Rosenberg}. This ferromagnetic order is also too weak to be detected by MOKE, perhaps because the photon wavelength used (1.8 ev) is far above the band gap of the bulk semiconductor (0.3 ev); magneto-optical effects in semiconductors are well known to be resonantly enhanced near optical transitions and can be much weaker under non-resonant conditions. The absence of the AHE could simply arise because it is masked by the ordinary Hall effect from the parallel bulk channel which dominates the transport. 

Having provided evidence for ferromagnetism in these \MBS~thin films, we now discuss the influence of this ferromagnetism on the MC in a magnetic field applied perpendicular to the sample plane (along the $c$ axis). Figures 6a - d show the longitudinal MC at various temperatures for four \MBS~thin film devices. If we focus on their behavior at low fields, a generic pattern emerges in all the samples: at high temperatures, all the samples show a negative MC which crosses over to a positive MC as the temperature is lowered. The cross-over coincides with the onset of ferromagnetic hysteresis in the MC. In contrast, the MC behavior of undoped \BS~is always dominated by a negative MC that has a characteristic cusp-like form at low fields, originating from weak anti-localization, and a (classical) parabolic or linear form at high fields.\cite{Richardella2010,Wang2011,MLiu2011} This contrast is apparent in Fig.\ 6e, where the MC response is depicted for samples of various Mn concentrations.

We now discuss an interpretation of these data based upon diagrammatic calculations\cite{Lu2011,Lu2011b} for quantum corrections to transport in \BS. In samples such as the ones studied here, the bulk states dominate the conductivity because of the large Fermi energy: a simple estimate using the surface state energy dispersion and a Fermi energy 300 meV above the Dirac point shows that the surface carrier density ($\sim1.3\times10^{13}$ cm$^{-2}$) is about an order of magnitude smaller than that of the bulk in our samples. Thus, an important question to address is the coexisting surface and bulk conduction in our samples. The classical contribution to the MC from bulk channels is well known to result in a parabolic positive magnetoresistance or a negative MC. According to diagrammatic calculations\cite{Lu2011b}, the quantum corrections to the conductivity of the lowest 2D bulk quantum well states in \BS~ are purely in the orthogonal class and can only result in weak localization (positive MC). In contrast, similar calculations for the Dirac surface state show that the quantum corrections to the conductivity are in the symplectic class and result in weak anti-localization (negative MC). The calculations also show that the opening of a gap at the Dirac point leads to a change in the quantum corrections away from the symplectic class, causing a transition from negative MC to positive MC with the onset of ferromagnetism\cite{Lu2011}. This transition is tunable by the relative size of the gap $\Delta$ and the Fermi energy $E_F$ measured from the Dirac point. In the limit of small $\Delta/2E_F$, the MC is dominated by weak anti-localization and is negative. In the limit of large $\Delta/2E_F$, the MC mainly originates in weak localization and is positive. At intermediate gap sizes, at low fields a relatively flat classical MC could be expected. While the large number of free parameters in the theoretical expressions of ref. 18 precludes meaningful fitting of our experimental data, it is still useful to compare the measured MC with simulations obtained from the theory. The MC is given by:\cite{Lu2011}
\begin{equation}
\Delta\sigma(B)=\sum_{i=0,1} \frac{\alpha_i e^2}{\pi h}\left[\Psi\left(\frac{l^2_B}{l^2_\phi}+\frac{l^2_B}{l^2_i}+\frac{1}{2}\right)-\ln \left(\frac{l^2_B}{l^2_\phi}+\frac{l^2_B}{l^2_i}\right)\right],
\end{equation}
where $\alpha_0$ and $\alpha_1$ are pre-factors for weak localization and weak anti-localization respectively, $\Psi(x)$ is the digamma function, $l_B$ is the magnetic length, $l_{\phi}$ is the phase coherence length, and $l_0$, $l_1$ are corrections to $l_{\phi}$. Using reasonable parameters ($l_{\phi}=300$ nm and magnetic scattering length $l_m=1000$ nm) for the simulation, the MC in Fig.\ 6f shows a cross-over from weak anti-localization to weak localization as we adjust $\Delta/2E_F$ from 0 (gapless surface state) to 0.99 (fully gapped surface state). This interpretation ascribes all weak anti-localization signatures in ungapped topological insulator thin films to surface state transport, regardless of the presence of a parallel bulk transport channel. We caution that our films are not thin enough to exclude the presence of quantum corrections to the MC from 3D bulk transport as well. These quantum corrections could very well result in a negative MC depending on the specific physical parameters \cite{Hikami1980}, but we are unaware of a detailed calculation for \BS. 
  
Applying this picture to our experimental observations leads to a consistent explanation. The observation of a negative MC at high temperatures ($T \gtrsim 6$K) or at high fields is most likely due to classical contributions from 2D/3D bulk states. The observed negative MC in the \MBS~samples is not characterized by the typical ``sharp cusp'' seen in weak anti-localization of \BS~samples, though some sign of it can be seen in the magnetoresistance above the bulk $T_C$ (Fig. 3 (c)).  Since ARPES and XMCD clearly show surface ferromagnetic order and a gap at high temperatures, the high temperature curves are likely to correspond to the low to intermediate  $\Delta/2E_F$ regime, similar to the relatively flat 0.2 ratio curve in Figure 6f. As the temperature is lowered ($T \lesssim 6$ K), besides the presence of the out-of-plane high temperature surface ferromagnetic phase, a limited near-surface bulk region also undergoes a transition to the ferromagnetic phase which is largely in-plane but also has a small out-of-plane component (Fig. 2). This in-plane ferromagnetism would not be expected to have much effect on the gap except that the coercive field is small enough that the application of the out-of-plane external measurement field is enough to rotate its orientation out-of-plane.  At a field ($\approx 0.2 T$) near where the hysteresis loop closes, indicating the moments are uniformly pointing out-of-plane, there is a transition to a region of strong positive MC consistent with a larger opening of the gap and an increased $\Delta/2E_F$ ratio pushing the system into weak localization.\cite{Lu2011} This interpretation is also consistent with the observed absence of MC at low perpendicular fields: in this regime, the magnetization is still mainly in-plane and does not contribute much to the further opening of the surface state gap, while the disordered out-of-plane moments that do exist could be expected to suppress quantum corrections by introducing random phases in the backscattered loops.  However, once the perpendicular field is large enough, the magnetization begins to be dominated by a uniform out-of-plane component, resulting in the positive MC before eventually turning over again due to the classical negative MC at high fields (see for example Figs. 6 (a) and (b)). The dependence on the Mn doping can also be understood. For the lightly doped samples (Bi/Mn $\geqslant12.5$), the carrier concentration is about $4\times10^{19}$ cm$^{-3}$. The increase in $\Delta$ with a relatively low $E_F$ gives a larger $\Delta/2E_F$, hence the positive MC gets enhanced. As we further increase the Mn concentration (Bi/Mn $\leqslant10.3$), any further increase in $\Delta$ is offset by a considerable rise in the Fermi energy as additional Mn n-dopes the material (the carrier concentration increases to $n\sim2\times10^{20}$ cm$^{-3}$). As a result, the change in positive MC is relatively small compared to the lightly-doped samples.

\section{ARPES measurements}
As we mentioned in the introduction, a detailed study and analysis of ARPES data (including spin-resolved measurements) has been reported elsewhere\cite{Xu_NP} and discusses the complex spin configuration of the surface states and its time reversal breaking nature as a result of magnetic doping. For the sake of completeness and to discuss the relationship between the magnetometry, magneto-transport and spin-resolved ARPES, we briefly describe the salient features of these surface-sensitive ARPES measurements. These experiments were carried out using Se-passivated \MBS~samples (Bi/Mn $=15$) which were prepared identically to those used in STM experiments. Again, the capped thin films were heated up carefully under ultrahigh vacuum ($<1\times10^{-9}$ Torr) to remove the Se capping layer. ARPES measurements were performed at 20 K on beamline 10.0.1 at the Advanced Light Source in Lawrence Berkeley National Laboratory. Typical energy and momentum resolution is better than 10 meV and $1 \%$ of surface Brillouin zone, respectively. ARPES core level spectroscopy data (Fig.\ 7a) show that only the Se core level was observed before the decapping process (blue curve), whereas both Se and Bi peaks appeared after the decapping (red curve). Figures 7 (b) and 7 (c) allow a clear comparison between ARPES measured dispersion mappings of undoped \BS~and \MBS~thin films: while a bright and intact Dirac point is observed for the undoped \BS~sample, the energy spectra weight of the Dirac node of the \MBS~samples is strongly suppressed. This comparison indicates that when time reversal symmetry is broken by introducing magnetic impurities into the system, the degeneracy of the electronic states at the Dirac node is lifted. Nevertheless, the suppressed intensity at the Dirac point does not go to zero. The absence of a fully developed gap may be attributed to the inhomogeneous Mn concentration on the surface, causing a spatially dependent magnetic field from the Mn atoms or surface clusters. 

Note that these surface-sensitive ARPES measurements were carried out at temperature $T = 20$ K, which is higher than the Curie temperature observed by magnetization and transport measurements. However, since the topological surface states are reported to localize mainly within the top quintuple layer (also within 1 nm) \cite{Xue Nature physics QL}, the electronic structure of the surface states is most relevant to the magnetic properties of the surface. The electronic structure within 1 nm of the terminated surface has multiple components that comes with spin physics: spin polarized topological surface states whose texture is modified by Mn-moments and bulk bands that are bent near the surface which also develop spin polarization through Rashba-type effect near the surface. In addition, surface magnetic properties can significantly differ from the bulk due to the strongly inhomogeneous Mn concentration profile along the out-of-plane $z$ direction and the possible surface magnetic ordering metallically mediated by topological surface conduction electrons with high mobility. The spin-resolved ARPES and XMCD measurements show an out-of-plane magnetic component on the very surface (possibly within 1 nm) region of the films under UHV conditions. Therefore, it is probably necessary to consider the effects and influence of both the in-plane and the out-of-plane magnetic moment components as well as bulk moments of these \MBS~ films to understand the full details of magnetism probed in transport and magnetometry. The Dirac point spectral weight suppression seen in Fig. 7 (c) originates from multiple physical mechanisms: first, the out-of-plane magnetic component opens up a time reversal symmetry breaking gap at the Dirac point, which leads to Dirac point spectral weight suppression under the inhomogeneous Mn distribution on the film surface; second, the in-plane magnetic component also leads to spectral weight suppression (without opening a gap at the Dirac point). An in-plane magnetization along a fixed momentum-space direction shifts the Dirac point away from the $\bar{\Gamma}$ point along the magnetization direction. At $T = 20$ K, the in-plane magnetic moment component on the surface is randomly distributed rather than aligned along a fixed direction. Such random in-plane magnetic component consequently leads to a substantial broadening of the surface states along the momentum axis. It has been shown by several ARPES experiments \cite{Momentum broadening, false gap arXiv} that substantial momentum broadening of a Dirac-like band leads to a spectral weight suppression (a ``gap''-like feature) at the Dirac point. Additionally, such observations are also understood by STM measurements \cite{spatial fluctuation} in the real-space point of view by the momentum fluctuation of the surface Dirac fermion in real space. Our low temperature transport measurements on \MBS~ thin films at $T < 6$ K are likely related to the combined effect and influence of the in-plane and out-of-plane (purely surface origin) magnetic component (the momentum broadening of the surface states) on the shown ARPES measurements. A complete picture will require a detailed understanding of the interplay between the bulk moments, surface moments and the role of additional Rashba bands near the surface. Due to the inherent surface-sensitive nature of ARPES, currently it is not possible to probe bulk magnetism with ARPES. 

{\bf Acknowledgement}

We thank S. -Q. Shen and H. -Z. Lu for valuable comments and discussion. This work was supported by DARPA (N66001-11-1-4110). We also acknowledge partial support from ONR (N00014-12-1-0116 and -0117), the Penn State Center for Nanoscale Science under the MRSEC program (NSF grant DMR-0820404). This publication was also supported by the Pennsylvania State University Materials Research Institute Nanofabrication Lab and the National Science Foundation Cooperative Agreement No.\ ECS-0335765.

\newpage
\begin{figure}
\includegraphics[width=4in]{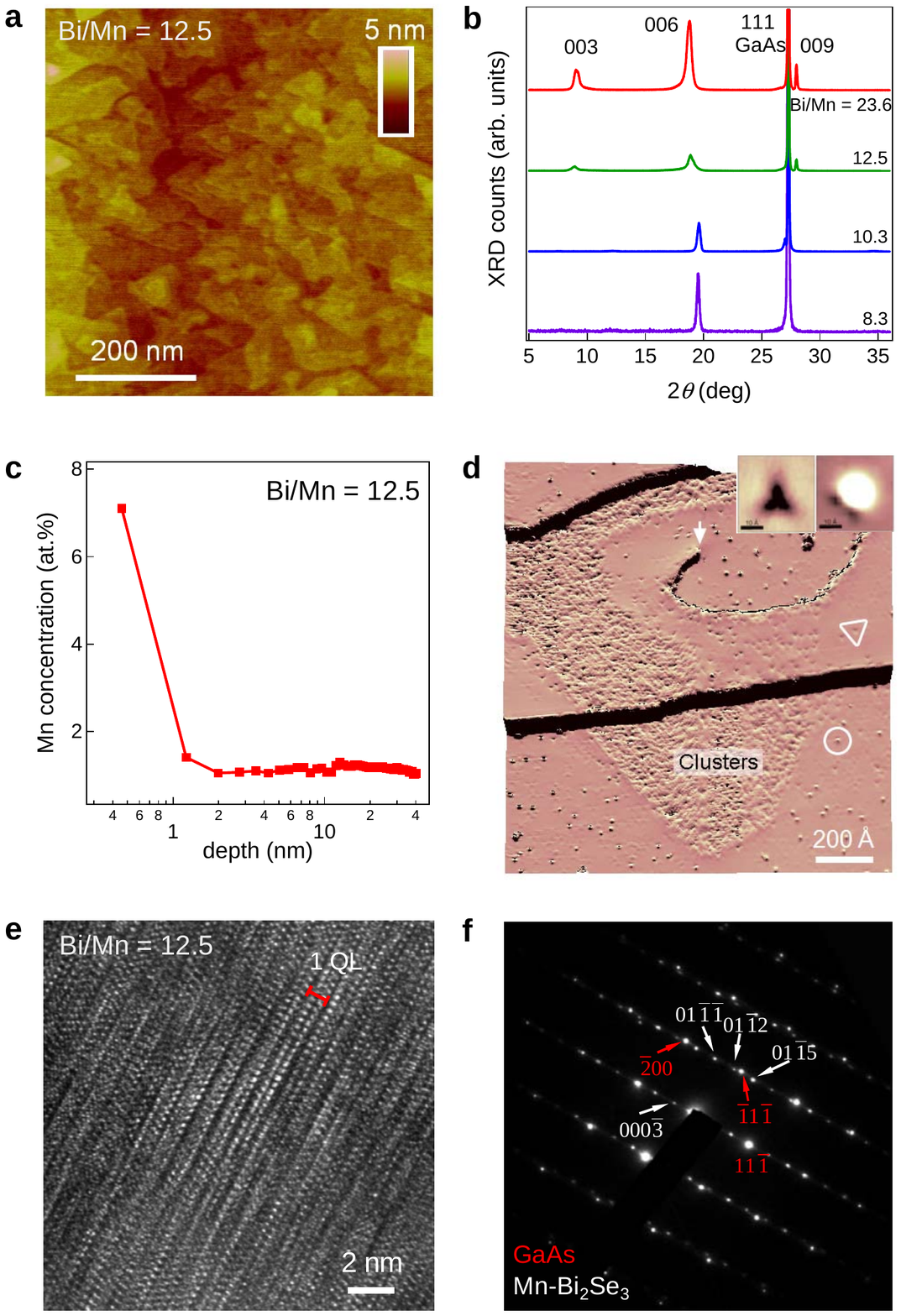} 
\caption{(Color online) (a) AFM image of a \MBS~thin film. (b) XRD from four \MBS~thin films. (c) SIMS depth profile showing a higher Mn concentration near the surface. (d) STM topographic image ($150\times150$ nm$^2$, 400mV, 40pA) of the surface of a \MBS~thin film after desorbing the Se capping layer. Substitutional Mn (triangle) clusters around the atomic step edges (dark line). Mn diffused onto the surface as adatoms is indicated by the circle and the screw dislocation is indicated by the arrow. The insets zoom in on a substitutional Mn atom and a Mn adatom (scale bar: 1 nm). (e) Cross-sectional HRTEM image of a \MBS~thin film. (f) Selected area electron diffraction pattern from the same \MBS~thin film in panel e.}
\label{Fig1}
\end{figure}

\newpage
\begin{figure}
\includegraphics[width=6in]{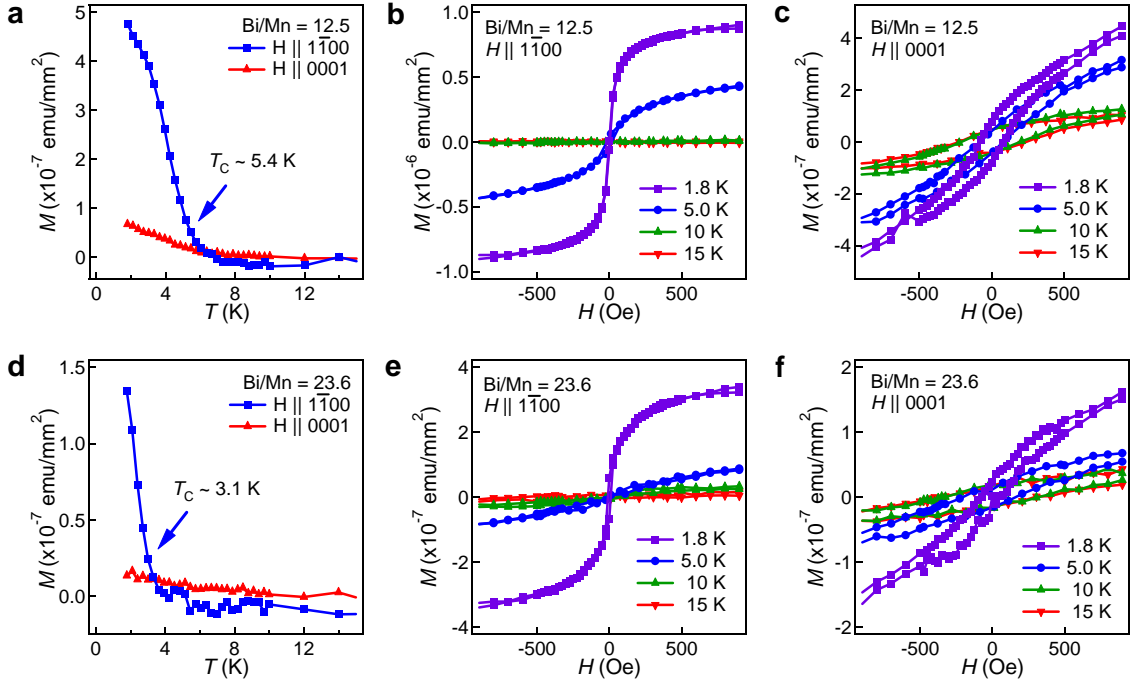} 
\caption{(Color online) (a) Temperature dependent magnetization of the Bi/Mn = 12.5 sample in field along $\langle1\bar{1}00\rangle$ direction and the $c$ axis. The arrow indicates the Curie temperature. (b) Magnetization vs. field plots of the Bi/Mn = 12.5 sample in field along $\langle1\bar{1}00\rangle$ direction at different temperatures. (c) Magnetization vs. field plots of the Bi/Mn = 12.5 sample in field along the $c$ axis at different temperatures. (d) Temperature dependent magnetization of the Bi/Mn = 23.6 sample in field along $\langle1\bar{1}00\rangle$ direction and the $c$ axis. The arrow indicates the Curie temperature. (e) Magnetization vs. field plots of the Bi/Mn = 23.6 sample in field along $\langle1\bar{1}00\rangle$ direction at different temperatures. (f) Magnetization vs. field plots of the Bi/Mn = 23.6 sample in field along the $c$ axis at different temperatures.}
\label{Fig2}
\end{figure}

\newpage
\begin{figure}
\includegraphics[width=5in]{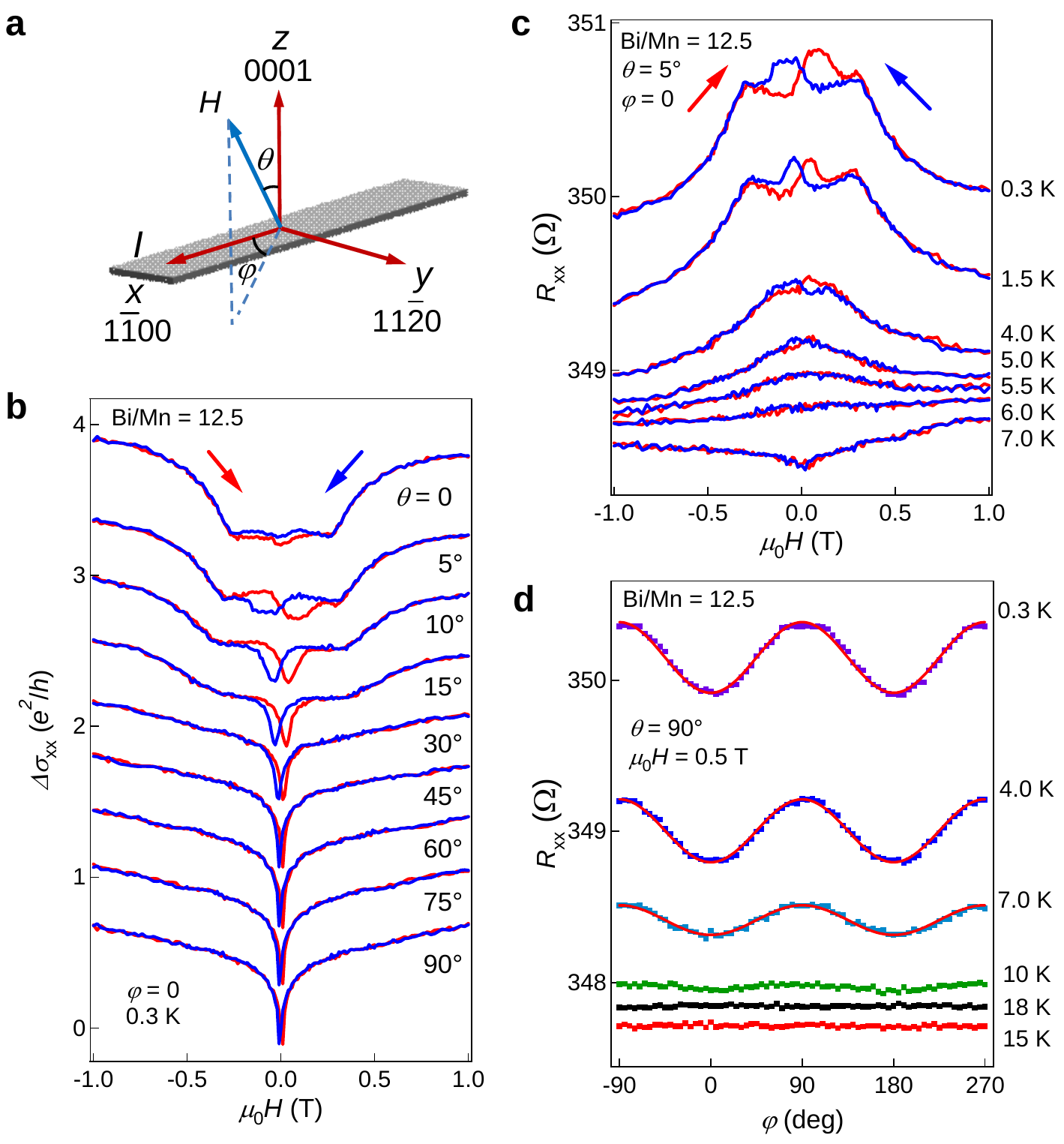} 
\caption{(Color online) (a) Measurement geometry: the \MBS~Hall bar lies in $xy$ plane with the excitation current flowing along $x$ axis. The orientation of the field $H$ is denoted by the polar angle $\theta$ and azimuth angle $\varphi$. (b) $\Delta\sigma_{xx}$ versus $H$ plots in different field orientations in $xz$ plane. The arrows indicate the field sweep directions and the curves at different field orientations are offset for clarity with respect to the $\theta=90^{\circ}$ curve. (c) $R_{xx}$ versus $H$ plots in field sweeps ($\theta=5^{\circ}$ and $\varphi=0$) at different temperatures. The hysteresis disappears when weak anti-localization behavior starts to dominate at $T\sim 5.5$ K. (d) $R_{xx}$ versus $H$ plots in $xy$ plane at different temperatures. The red lines are fits to equation (1). }
\label{Fig3}
\end{figure}

\newpage
\begin{figure}
\includegraphics[width=6in]{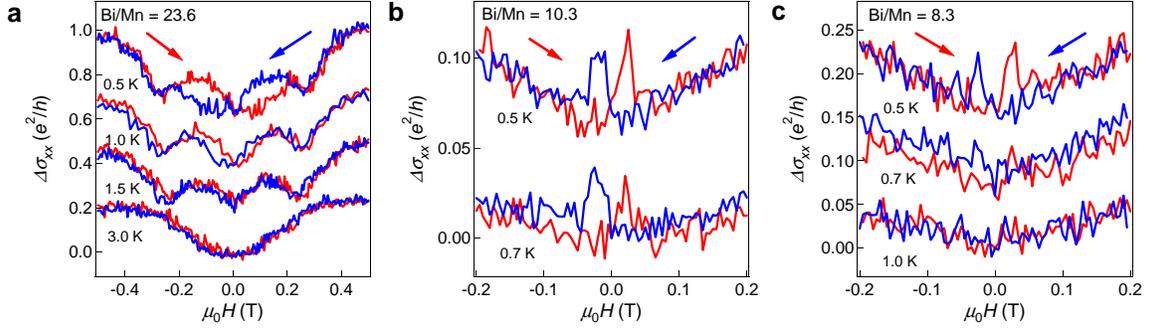} 
\caption{(Color online) MC of additional three \MBS~samples in perpendicular field ($H\parallel c$ axis). (a) MC hysteresis is observed in the most lightly doped (Bi/Mn = 23.6) sample. Hysteretic MC jumps are observed in the heavily doped samples (Bi/Mn $\leqslant10.3$), shown in panel (b) and (c). The arrows indicate the field sweep directions and the curves are shifted with respect to the curve measured at the highest temperature for clarity.}
\label{Fig4}
\end{figure}

\newpage
\begin{figure}
\includegraphics[width=2in]{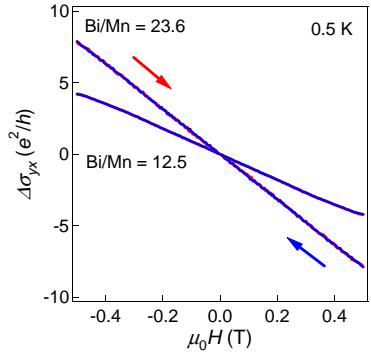} 
\caption{(Color online) Hall conductivity of samples Bi/Mn = 23.6 and 12.5 at 0.5 K. The arrows indicate the field sweep directions.}
\label{Fig5}
\end{figure}

\newpage
\begin{figure}
\includegraphics[width=5in]{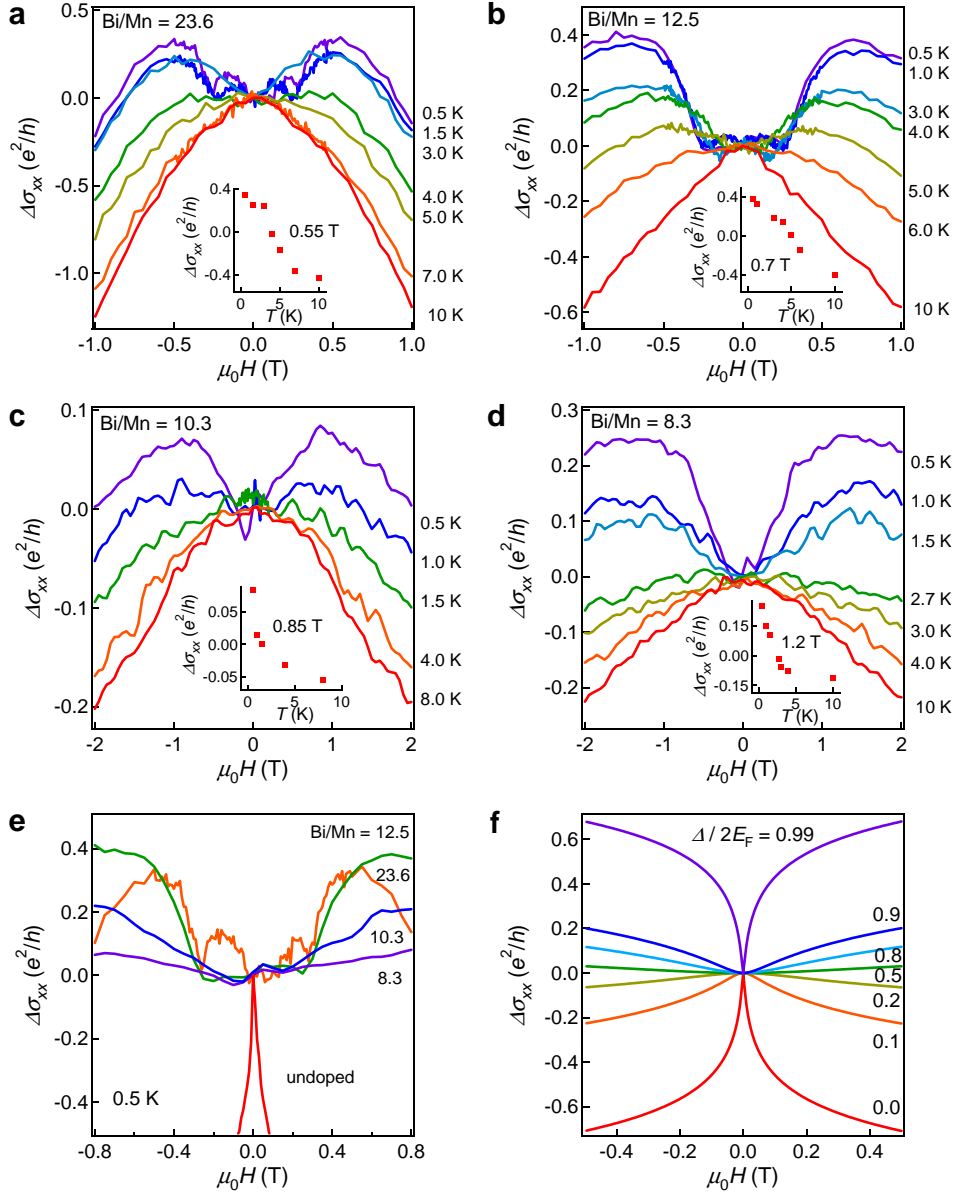} 
\caption{(Color online) MC of \MBS~in field perpendicular to the sample plane. (a) MC of the most lightly doped sample (Bi/Mn = 23.6) shows a cross-over from weak anti-localization to weak localization at $T\sim3.0$ K, consistent with the $T_C$ obtained from SQUID measurements. (b) MC of the Bi/Mn = 12.5 sample shows larger positive MC, which survies up to $\sim5$ K. (c) MC of the Bi/Mn = 10.3 sample. (d) MC of the most highly doped sample (Bi/Mn = 8.3). The insets in panel (a) - (d) plot temperature dependent $\Delta\sigma_{xx}$ at fixed perpendicular magnetic field. (e) MC of four \MBS~and one undoped \BS~sample. (f) Simulation using euqation (2) showing cross-over from weak anti-localization to weak localization as parameter $\Delta/2E_F$ changes.}
\label{Fig6}
\end{figure}

\newpage
\begin{figure}
\includegraphics[width=3in]{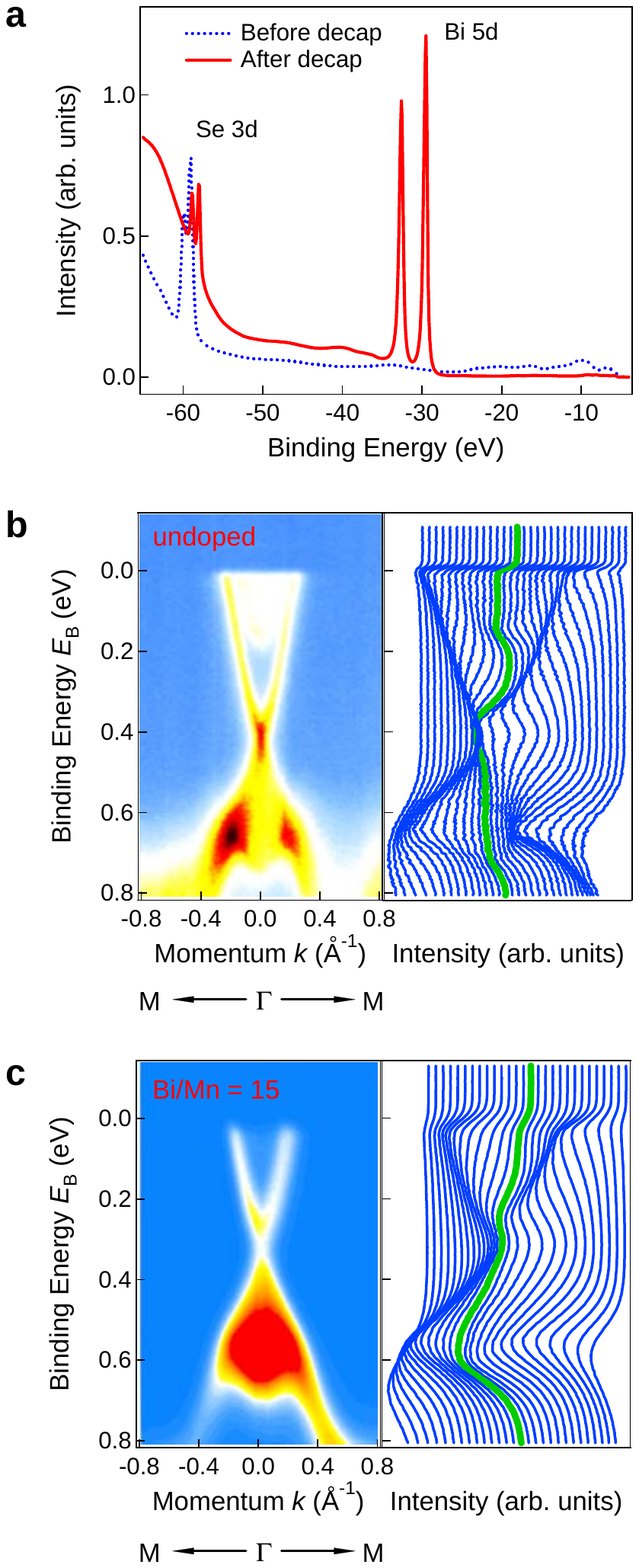} 
\caption{(Color online) (a) Core level spectroscopies on \MBS~thin film before and after the decapping procedure. (b) High resolution ARPES dispersion mapping and corresponding energy dispersion curves (EDCs) of an undoped \BS~thin film along high symmetry $M - \Gamma - M$ direction. The EDC that crosses the $\Gamma$-point is highlighted in green color. (c) High resolution ARPES dispersion mapping and corresponding EDCs of Bi/Mn $=15$ \MBS~thin film along high symmetry $M - \Gamma - M$ direction. The EDC that crosses the $\Gamma$-point is highlighted in green.}
\label{Fig7}
\end{figure}


\newpage
\begin{thebibliography}{99} 
\bibitem{Moore_2010} J. E. Moore, {\it Nature} {\bf 464}, 194 (2010).
\bibitem{Qi2010} M. Z. Hasan and C. L. Kane, {\it Rev. Mod. Phys.} {\bf 82}, 3045 (2010); X. -L. Qi and S. -C. Zhang, {\it ibid} {\bf 83}, 1057 (2011).
\bibitem {Qi2008}X. -L. Qi, T. L.  Hughes, and S. -C.  Zhang, {\it Nat. Phys.} {\bf 4}, 273 (2008).
\bibitem{Qi2008b}X. -L. Qi, T. L.  Hughes, and S. -C.  Zhang, {\it Phys. Rev. B} {\bf 78}, 195424 (2008).
\bibitem{Liu2009}Q. Liu, C. -X. Liu, C. Xu, X. -L. Qi, and S. -C. Zhang, {\it Phys. Rev. Lett.} {\bf 102}, 156603 (2009).
\bibitem{Yu2010}R. Yu, {\it et. al.} {\it Science} {\bf 329}, 61 (2010).
\bibitem{Qi2009}X. -L. Qi, R. Li, J. Zang, and S. -C. Zhang, {\it Science} {\bf 323}, 1184 (2009).
\bibitem{Wray2010} L. A. Wray, {\it et. al.} {\it Nat. Phys.} {\bf 7}, 32 (2010).
\bibitem{Chen2010} Y. L. Chen, {\it et. al.} {\it Science} {\bf 329}, 659 (2010).
\bibitem{Hor2010}Y. S. Hor, {\it et al.} {\it Phys. Rev. B} {\bf 81}, 195203 (2010).
\bibitem{Checkelsky2012} J. G. Checkelsky, Jianting Ye, Yoshinori Onose, Yoshihiro Iwasa and Yoshinori Tokura, {\it Nat. Phys.} {\bf 8}, 729(2012).
\bibitem{Zhou2005} Z. H. Zhou, Y. -J. Chien, and C. Uher. {\it Appl. Phys. Lett.} {\bf 87}, 112503 (2005).
\bibitem{Zhou2006} Z. H. Zhou, Y. -J. Chien, and C. Uher. {\it Phys. Rev. B} {\bf 74}, 224418 (2006).
\bibitem{Chang2011}C. -Z. Chang, {\it et. al.} arxiv:1108.4754.
\bibitem{Haazen2012}P. P. J. Haazen, {\it et. al.} {\it Appl. Phys. Lett.} {\bf 100}, 082404 (2012).
\bibitem{Xu_NP}S.-Y. Xu, {\it et al.} {\it Nat. Phys.} {\bf 8}, 616 (2012).
\bibitem{Rosenberg}G. Rosenberg and M. Franz, {\it Phys. Rev. B} {\bf 85,} 195119 (2012).
\bibitem{Lu2011}H. -Z. Lu, J. Shi, and S. -Q.  Shen, {\it Phys. Rev. Lett.} {\bf 107,} 076801 (2011).
\bibitem{Liu2011}M. Liu, {\it et al.} {\it Phys. Rev. Lett.} {\bf 108,} 036805 (2012).
\bibitem{DietlNM} S. Kuroda, {\it et. al.} {\it Nat. Mater.} {\bf 6}, 440 - 446 (2007).
\bibitem{Pan1957}J. P. Pan, {\it Solid State Physics, edited by Seitz, F. \& Turnbull, D.} Vol. 5, p1-96 (Academic, New York, 1957).
\bibitem{Wang2011b}J. Wang, {\it et. al.} arxiv:1108.1465.
\bibitem{Kilanski2011}L. Kilanski, {\it et. al.} {\it Solid State Commun.} {\bf 151,} 870 (2011).
\bibitem{Bardelben} J. Von Bardelben {\it et al.} (unpublished).
\bibitem{Richardella2010}A. Richardella, {\it et. al.} {\it App. Phys. Lett.} {\bf 97,} 262104 (2010).
\bibitem{Wang2011}J. Wang, {\it et. al.} {\it Phys. Rev. B} {\bf 83,} 245438 (2011).
\bibitem{MLiu2011}M. Liu, {\it et. al.} {\it Phys. Rev. B} {\bf 83,} 165440 (2011).
\bibitem{Lu2011b} H.-Z. Lu and S. -Q. Shen, {\it Phys. Rev. B} {\bf 84}, 125138 (2011).
\bibitem{Hikami1980} S. Hikami, A. I. Larkin and Y. Nagaoka, {\it Prog. Theor. Phys.} {\bf 63}, 707 (1980).
\bibitem{Xue Nature physics QL}Y. Zhang, {\it et. al.} {\it Nat. Phys.} {\bf 6,} 584 (2010).
\bibitem{spatial fluctuation} H. Beidenkopf, \textit{et al} \textit{Nat. Phys.} $\mathbf{7}$, 939-943 (2011).
\bibitem{Momentum broadening} L. A. Wray, \textit{et al},  arXiv:1206.1087.
\bibitem{false gap arXiv} S.-Y. Xu, \textit{et al},  arXiv:1206.0278.
\end{thebibliography}
\end{document}